\documentclass[12pt,preprint]{aastex}

\shorttitle{Timing Features of XTE J1807$-$294}
\shortauthors{Zhang F. et al.}

\begin{document}

\title{Timing Features of the Accretion--driven Millisecond X-Ray Pulsar XTE J1807--294 in 2003 March Outburst}

\author{Fan Zhang\altaffilmark{1}, Jinlu Qu\altaffilmark{1},
 C. M. Zhang\altaffilmark{2}, W. Chen\altaffilmark{1}, T. P. Li\altaffilmark{1,3}}

\altaffiltext{1}{Key Laboratory of Particle Astrophysics,
Institute of High Energy Physics, CAS, Beijing 100049, PR
China}\email{zhangfan@mail.ihep.ac.cn,qujl@mail.ihep.ac.cn}

\altaffiltext{2}{National Astronomical Observatories, Chinese
Academy of Sciences, Beijing 100012, PR China}
\email{zhangcm@bao.ac.cn}

\altaffiltext{3}{Dept. of Physics and Center for Astrophysics,
Tsinghua University, Beijing 100084, PR China}
\email{litp@mail.ihep.ac.cn}

 \begin{abstract}

In order to probe the activity of the inner disk flow and its
effect on the neutron star surface emissions, we carried out the
timing analysis of the {\it Rossi X-Ray Timing Explorer}
observations of the millisecond X-ray pulsar XTE J1807--294,
focusing on its correlated behaviors in X-ray intensities,
hardness ratios, pulse profiles and power density spectra. The
source was observed to have a serial of  broad ``puny" flares on a
timescale of hours to days on the top of a decaying outburst in
March 2003. In the flares, the spectra are softened and the pulse
profiles become more sinusoidal.  The frequency of kilohertz
quasi-periodic oscillation (kHz QPO) is found to be positively
related to the X-ray count rate in the flares. These features
observed in the flares could be due to the
 accreting flow inhomogeneities. It is noticed that the fractional pulse
amplitude increases with the flare intensities in a range of $\sim
2\%-14\%$, comparable to those observed in the thermonuclear
bursts of the millisecond X-ray pulsar XTE J1814--338, whereas
it remains at about $6.5\%$ in the normal state. Such a
significant variation of the pulse profile in the ``puny" flares
may reflect the changes of physical parameters in the inner disk accretion region.
Furthermore, we noticed an overall positive correlation
between the kHz QPO frequency and the fractional pulse amplitude,
which could be the first evidence representing that the
neutron-star surface emission properties are very sensitive to the
disk flow inhomogeneities. This effect should be cautiously
considered in the burst oscillation studies.

\end{abstract}


\section{INTRODUCTION}

Neutron stars in low-mass X-ray binaries (LMXBs) accrete matter
from a companion with a mass of less than $1\,\rm{M_{\sun}}$ via
an accretion disk. In many models, the Keplerian accretion disk is
supposed to be terminated at an inner radius $r_{\rm in}$ of a few
Schwarzschild radii by e.g. relativistic effects, radiation drag,
or neutron-star magnetosphere-disk interactions (see van der Klis
2004 for a review). The motion of the inner disk flow and the
geometry of the innermost disk are still uncertain. The
observations of the correlated behaviors in X-ray spectral and
timing variabilities in LMXBs have provided the important probes
of the accretion-flow dynamics and the stellar-disk interactions.

Kilohertz quasi-periodic oscillations (kHz QPOs, in the frequency
range of 200 Hz--1300 Hz) have been observed in more than 20
accreting neutron star LMXBs (see van der Klis 2000, 2004 for the
recent  reviews). Their frequencies are usually regarded to be
associated with the Keplerian orbital frequency at some preferred
radius related to $r_{\rm in}$. An important evidence for the
movement of the inner edge of the disk under different mass
accretion rate comes from the observation of a positive
correlation of frequency vs. count rate on timescales of hours to
days in some low-luminosity LMXBs, i.e. ``atoll" sources
\citep{mend99}. A similar correlation (or in some cases an
anti-correlation) has been also found in some high-luminosity
LMXBs, i.e. ``Z" sources, in the form of the correlations with
curve length $S_{\rm z}$ along the track traced in an X-ray
color-color diagram \citep{wijn97,yu01,homan02}. The effect of
stellar radiations on the disk movement has also been observed,
e.g. in 4U 1608--52, where \citet{yu02} found the
anti-correlations between the kHz QPO frequency and the X-ray
count rate associated with mHz QPO due to nuclear burning on the
neutron star surface.

Nearly coherent brightness oscillations have been discovered
during thermonuclear X-ray bursts in some LMXBs (see Strohmayer \&
Bildsten 2003 for a review). Recent observations of the accreting
millisecond pulsars, SAX J1808.4--3658 \citep{chak03} and XTE
J1814--338 \citep{mss03}, confirm that the burst oscillation
frequency is extremely close to the spin frequency. It supports
the interpretation of the burst oscillation in terms of a ``hot
spot" on the stellar surface \citep{chak03,SB03}, which
  triggers the efforts to investigate X-ray emission properties
on the neutron star  surface by folded pulse profiles
\citep{stro03,bhat05,watt05}. The best-fitting parameters of the
pulsed fraction and the first harmonic can not only determine the
compactness of neutron star, but also help to figure out how the
nuclear burning front (the hot spot) propagates on the stellar
surface \citep{bhat05,watt05}. However, in the burst-oscillation
studies, because of the rather limited photon statistics on the
short burst timescale, many similar bursts have to be added to
improve the statistics of the folded pulse profiles.
 The underlying problem is that these bursts
may occur in different mass accretion rates. Till now, we know
little about what effects  of  the accretion rate fluctuations can
make on  the  neutron-star surface emissions or material
depositions.

This situation can be changed with the discovery of the accreting
millisecond X-ray pulsars (MXPs, see Wijnands 2005 for a latest
review). Their detectable coherent pulsations allow us to get the
pulse profile with good statistics over a long observational
period when the energy spectrum (or hardness ratios) varies only
slightly. According to the detailed spectroscopic and
pulse-profile analysis on SAX J1808.4--3658, \citet{pout03}
constructed a model for X-ray emissions from a hot spot on the
surface/boundary layer of a rapidly rotating neutron star in the
hard island state. They used it to constrain the compactness of
the central star, the position of the emitting region and the size
of the hot spot. Their investigations shed a light on further
studies of the variations of the surface X-ray emission regions
during the evolution of the accretion disk.

Among the discovered 7 millisecond X-ray pulsars, XTE J1807--294
is the best candidate for us to investigate the impact of disk
flow activities on the neutron-star surface emission for four
reasons: (1) Besides  SAX J1808.4--3658, XTE J1807--294 is also a
source which has been reported to have twin kHz QPOs \footnote{In
completion of our paper, we noted that Linares et al. (2005) used
the same data of
 XTE J1807--294 as we  analyzed. Our results are similar
 in kHz QPOs and different in the low frequency ranges, which might be due to
  the different choices of the data segments and different data grouping.}
\citep{wijn03,wijn05,lina05}.
 (2) The binary parameters of it have
been calculated by \citet{camp03} and \citet{kirs04} based on the
{\it XMM-Newton} observation, so we are able to correct the photon
arrival times for the orbital motions in producing pulse profiles.
(3) The time-averaged energy spectrum of the source is found to be
dominated by an optically-thin Comptonized component
\citep{fala05}, similar to that of SAX J1808.4--3658
\citep{Gier02}, so that we can take the X-ray emission model of
SAX J1808.4--3658 as a direct reference. (4) There are no
thermonuclear bursts reported in this source; thus we can focus on
studying the effect of disk evolutions. In the left 6 MXPs, i.e.,
SAX J1808.4--3658, XTE J1751--305, XTE J0929--314, XTE J814--338,
IGR J00291$+$5934 and HETE J1900.1--2455 (see Wijnands 2005 for a
review), the above four conditions cannot be satisfied
simultaneously. With the knowledge mentioned above, many basic
features of XTE J1807--294 have been available as well, such as
the shortest orbital period of $\sim 40$ minutes and a relatively
slow spin frequency of $\sim$ 191 Hz \citep{mark03c}. In addition,
this source locates at  $5^{\circ}.7$ away from the Galactic
center, with the best known position reported by \citet{mark03c}
based on a Chandra observation. Assumed  a distance of 8 kpc, the
source luminosity dropped from $1.3\times
10^{37}\,\rm{erg\,s^{-1}}$ on February 28, 2003,  to $3.6\times
10^{36}\,\rm{erg\, s^{-1}}$ on March 22, 2003 \citep{fala05}.

We analyze the evolution of light curves, hardness ratios, pulse
profiles and power density spectra of XTE J1807--294, using {\it
Rossi X-ray Timing Explorer} ({\it RXTE}) observations from
February 27 to March 31, 2003. Our results firstly show that the
positive kHz QPO frequency-count rate correlation on timescales of
hours to days is related to some ``puny" flares,  originated from
disk flow inhomogeneities; the behaviors of the pulsed fraction
and the first harmonic in the flares are different from those in
the non-flares. However, in the whole investigated episode, there
is a positive correlation between the pulsed fraction and the kHz
QPO frequency, which could be the first evidence that neutron-star
surface emissions are very sensitive to the disk flow
inhomogeneities. Furthermore, we describe the {\it RXTE}
observations and data analysis method in Section 2;  the results
of timing analysis are presented in Section 3 and discussions of
the results  are shown in Section 4.  Finally, we  make a brief
conclusion of our studies in Section 5.

\section{{\it RXTE} OBSERVATIONS AND DATA ANALYSIS}

We study the data obtained by the Proportional Count Array (PCA)
on board {\it RXTE} during the March 2003 outburst decay of XTE
J1807--294, using FTOOLS 5.3.

To have an overview of the source evolution, we extract light
curves from the PCA standard 2 mode data (detector 2 only, all
layers), and subtract the background produced by `pcabackest'
version 3.0 with the ``CM" faint-source model. We define soft
color as the ratio of count rates in the $4.1-7.0\,\rm{keV}$ and
$2.0-4.1\,\rm{keV}$ bands and hard color as the ratio of count
rates in the $13.3-30.2\,\rm{keV}$ and $7.0-13.3\,\rm{keV}$ bands.
In selecting good time intervals for data analysis, we give a
condition of PCU2$_{-}$on.eq.1 in addition to the standard
conditions for faint sources. Then we extract the light curves of
PCU 0, 1, 3, 4 separately and compare them with the light curve of
PCU2 respectively. Finally, we find the time intervals when PCU 4
meet instrumental flares and delete such bad time intervals,
making sure the following analysis are not influenced by
instrumental problems.

For Fourier analysis, we construct pulse profiles and power
density spectra from the $\sim 0.95\,\mu$s time-resolution PCA
GoodXenon mode data (all available detectors, all layers),
correcting the photon arrival times for orbital motions of the
pulsar and the spacecraft. When dealing with the GoodXenon mode
data, we do not perform background subtraction or deadtime
correction.

For pulse-profile fitting we create folded light curves in 32
phase bins for every continuous event file. As a fit function of
the pulse profiles we adopt a two-component Fourier function of
the form
\begin{equation}
 I(\phi)=c_0(1+a_0\sin
[2\pi(\phi-\phi_0)]+a_1\sin [2\pi(\phi-\phi_1)/0.5])
\end{equation}
where $c_0$ is the normalized count rate of the pulse profile;
$a_0$ and $a_1$ are the fractional amplitude of the fundamental
and the first harmonic component, respectively; $\phi_0$ and
$\phi_1$ are the corresponding zero phase angles. We fit the pulse
profiles and determine the errors of the fit parameters by using
$\Delta \chi^2=2.7$. We estimate the real averaged count rate from
the $2-60\,\rm{keV}$ net light curve, produced by processing
Standard 2 mode data (detector 2, all layers) within the
corresponding intervals of every folded pulse profile.

We construct power density spectrum (PDS) per observation using
the data segments of 256 s and 1/4096 s time bins, so that the
lowest available frequency is 1/256 Hz and the Nyquist frequency
is 2048 Hz. We normalize the power spectra by the method of
\citet{miya91}, and subtract the Possion white noise. In order to
improve the statistics, we add the observations into groups when
these observations are adjacent in time and the power spectra
remain the same. The division of the groups can be found in Table
\ref{tab_observation} and are depicted in the top panel of Fig.
\ref{fig_outburst}. Most of our power spectra show a
power-law-like component in mHz--Hz frequency range, which could
be the very low frequency noise (VLFN). Furthermore, in our PDS,
some additional weak noises can be measured with good statistics
as shown in Section 3.3.

We use a power law plus multi--Lorentzian function
\citep{bpk02,vstr02} to fit the spectra. Before fitting the power
spectra, we remove frequency bins containing the pulse spike. To
visualize the characteristic frequencies ($\nu_{\rm max}$) of
noises or QPOs, we plot the power spectra in the power times
frequency representation ($\nu{\rm P}_{\nu}$), where ${\rm
P}_{\nu}$ is the power spectral density and $\nu$ is the Fourier
frequency. In this representation, the Lorentzian's maximum occurs
at $\nu_{\rm max}$ ($\nu_{\rm max} = \sqrt{\nu_0^2 + \Delta^2}$,
where $\nu_{\rm max}$ is the characteristic frequency, $\nu_0$ is
the centroid frequency and $\Delta$ is the half width and half
magnitude (HWHM) of the Lorentzian). We represent the Lorentzian
relative width by $Q$ defined as $\nu_0/2\Delta$, and the
root-mean-square fractional amplitude by $rms$.

\section{RESULTS}

\subsection{Broad ``Puny" Flares in 2003 March }

\begin{figure}[ht]
\epsscale{1.} \plotone{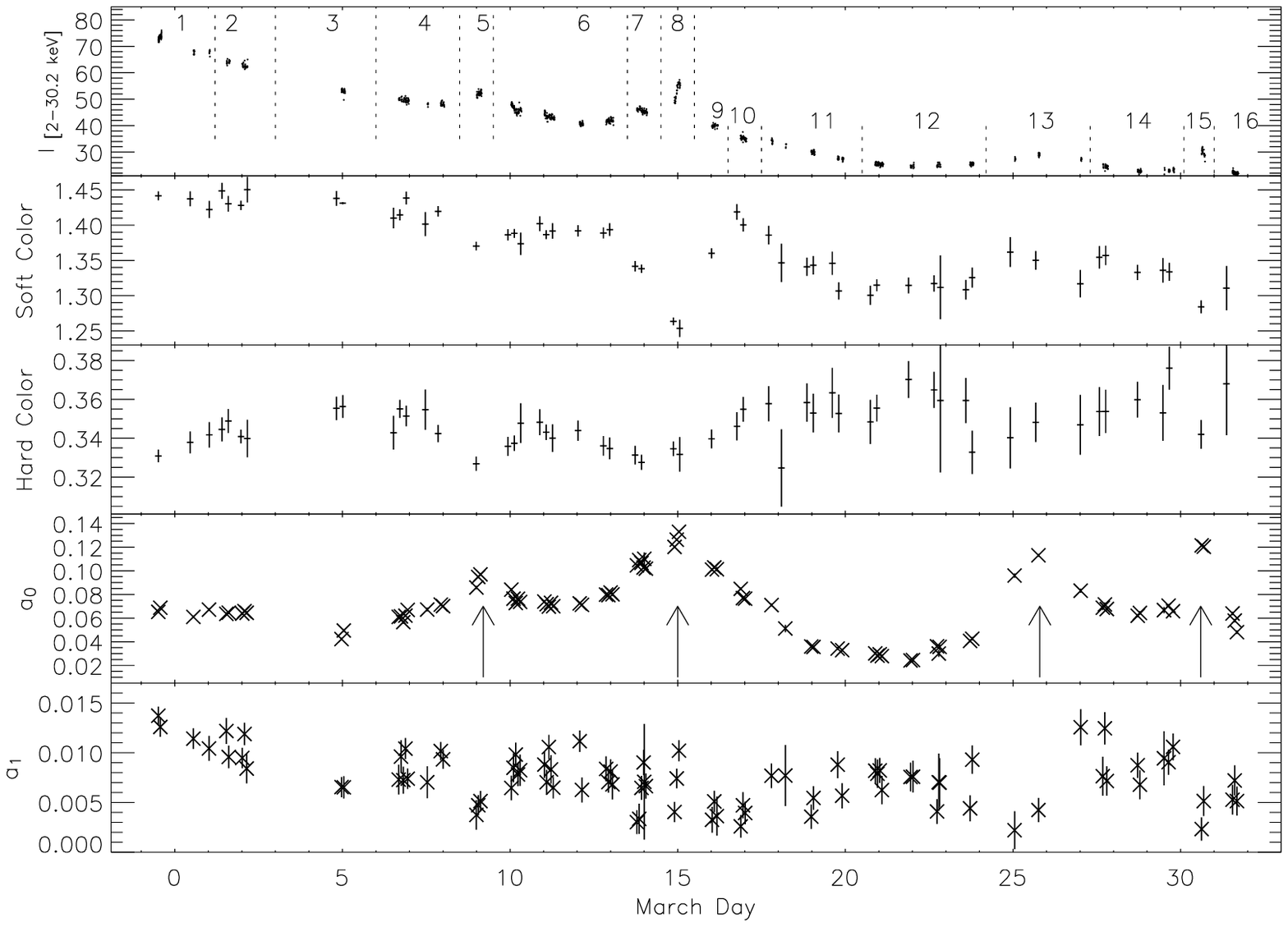} \caption{Evolution of the light
curve, soft/hard color, fractional amplitude of the pulsation
($a_0$) and the first harmonic component ($a_1$) of XTE J1807--294
in March, 2003. The light curve of $2.-30.2\,\rm{keV}$ is in unit
of $\rm{counts\, s^{-1}\,PCU^{-1}}$ (with time bins of 256 s); the
soft color is defined as $I_{[4.1-7.0\,{\rm
{keV}}]}/I_{[2.0-4.1\,{\rm {keV}}]}$, and the hard color is
defined as $I_{[13.3-30.2\,{\rm {keV}}]}/I_{[7.0-13.3\,{\rm
{keV}}]}$ (with time bins of 16482 s). The horizontal coordinate
stands for the date of the observations, where -1 and 0 denote the
days of February 27 and 28, respectively. The four arrows in the
fourth panel indicate the positions of the flares.
  }\label{fig_outburst}

\end{figure}

It has been found that the light curve of XTE J1807--294 drops
exponentially from February to the end of March of 2003
\citep{fala05} and then becomes flattened instead of being much
steepened as in SAX J1808.4-3658 \citep{Gilf98,Gier02} and XTE
J1751--305 \citep{GP05}. Figure \ref{fig_outburst} shows that in
the former exponentially decaying episode, at least four intensity
fluctuations occur, each lasting beyond several thousands of
seconds, with small X-ray count rate variations. To make
quantitative analysis, we fit the light curve by an exponential
plus multi-Gaussian model. The full width and half magnitude
(FWHM) of each Gaussian and the ratio of the magnitude of the
Gaussian to the exponential value at the same time are
($0.53\,\rm{day}$, $7.18/45.42$), ($0.95\,\rm{day}$,
$16.8/35.22$), ($1.67\,\rm{day}$, $7.53/20.75$),
($0.71\,\rm{day}$, $12.33/17.16$), respectively. The features of
long duration and low amplitude make these fluctuations different
from type I thermonuclear bursts and super nuclear outburst
\citep{SB02, SM02} on the neutron star surface. We thus call them
``puny" flares for convenience.
 In the following, it can be
noticed that these ``puny" flares are also featured by the
 enhanced soft X-ray emissions and increased pulsed fractions.

\begin{figure}[ht]
\epsscale{1.0} \plotone{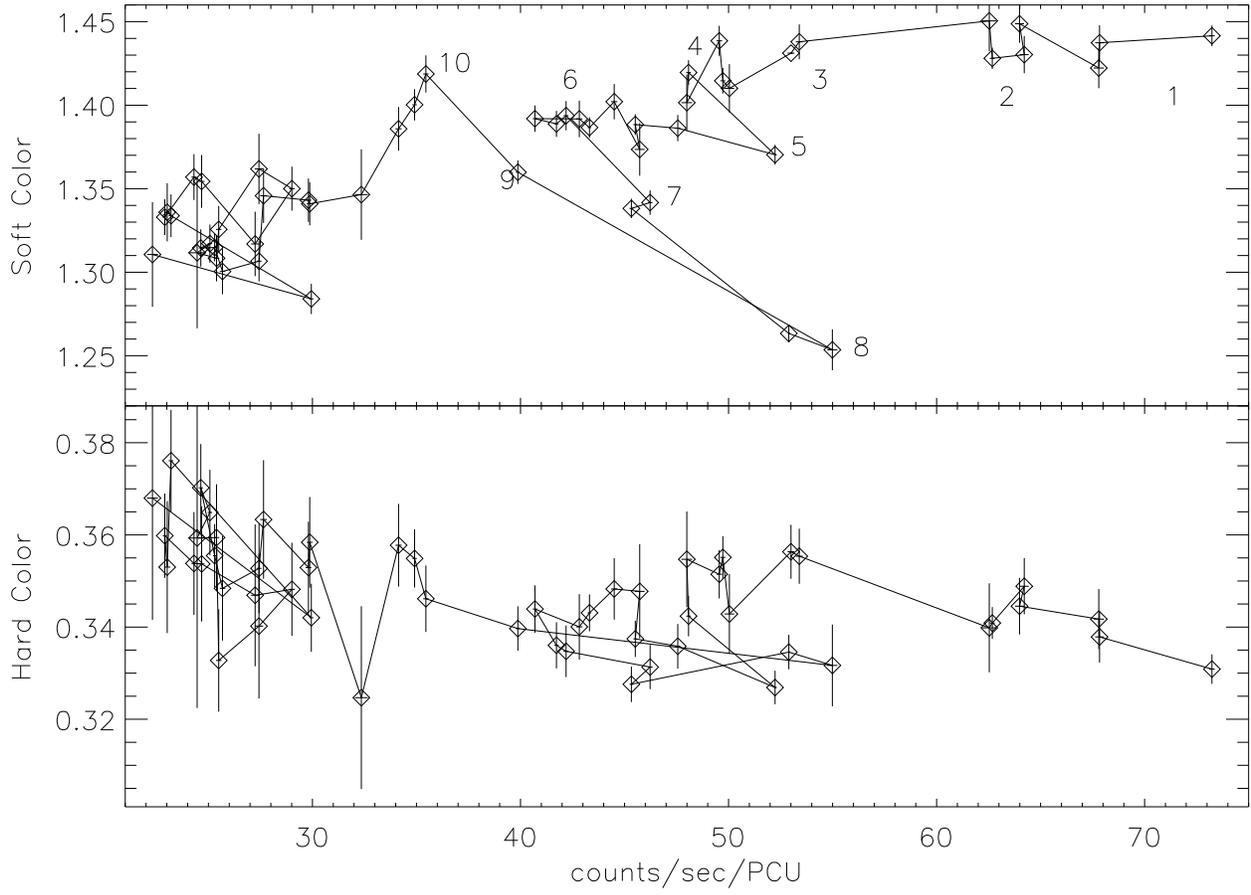} \caption{Hardness intensity
diagram of XTE J1807--294. The group numbers for every power
density spectra are noted in the top panel.}\label{fig_J1807_hid}

\end{figure}

The second panel of Fig. \ref{fig_outburst} shows that the soft
color drops in the flares, especially at the top of the largest
flare of Mach 14--15. By comparing the light curves in the four
energy bands ($2.0-4.1\,\rm{keV}$, $4.1-7.0\,\rm{keV}$,
$7.0-13.3\,\rm{keV}$ and $13.3-30.2\,\rm{keV}$), we find that this
is due to the stronger soft X-ray emissions in the band of
$2.0-4.1\,\rm{keV}$. The color-intensity diagram of Fig.
\ref{fig_J1807_hid} indicates that, except of the ``puny" flares,
the soft color decreases with the decay of the outburst, whereas
the hard color changes a little in the whole investigated episode.

We can deduce from the almost invariant hard color that the hard
spectral component changes only slightly in the whole episode. The
fitting results of the combined spectra of {\it XMM-Newton}, {\it
RXTE} and {\it INTEGRAL} observations \citep{fala05} present that
hard emissions from the hot and optically thin Comptonized plasma
dominate the energy spectra; the contribution of thermal emissions
from the accretion disk at February 28 and March 22 are so small
that they are almost negligible in PCA analysis (above 2.5 keV).
Therefore, the PCA X-ray emissions can be considered to be mainly
generated from the neutron star surface/boundary layer.

\subsection{Variations of Pulse Profiles}

We fit 86 pulse profiles with a two-component sine function
described in Section 2. The fitting is good in most cases, with
$\chi^2/\rm{dof}$ ranged in $32/59-80/59$. The situations of
$\chi^2/\rm{dof}>1.3$ only occur in three pulse profiles. We plot
the evolutions of the best-fitting parameters of $a_0$ and $a_1$
in the fourth and the fifth panel of Fig. \ref{fig_outburst},
respectively. It can be seen that $a_0$ is a rather unambiguous
indicator of the ``puny" flares than the X-ray intensities and
soft colors, not only because of its small errors, but also
because $a_0$ increases significantly at the top of the flares
despite how weak the flare is.

\begin{figure}[ht]
\epsscale{1.0} \plotone{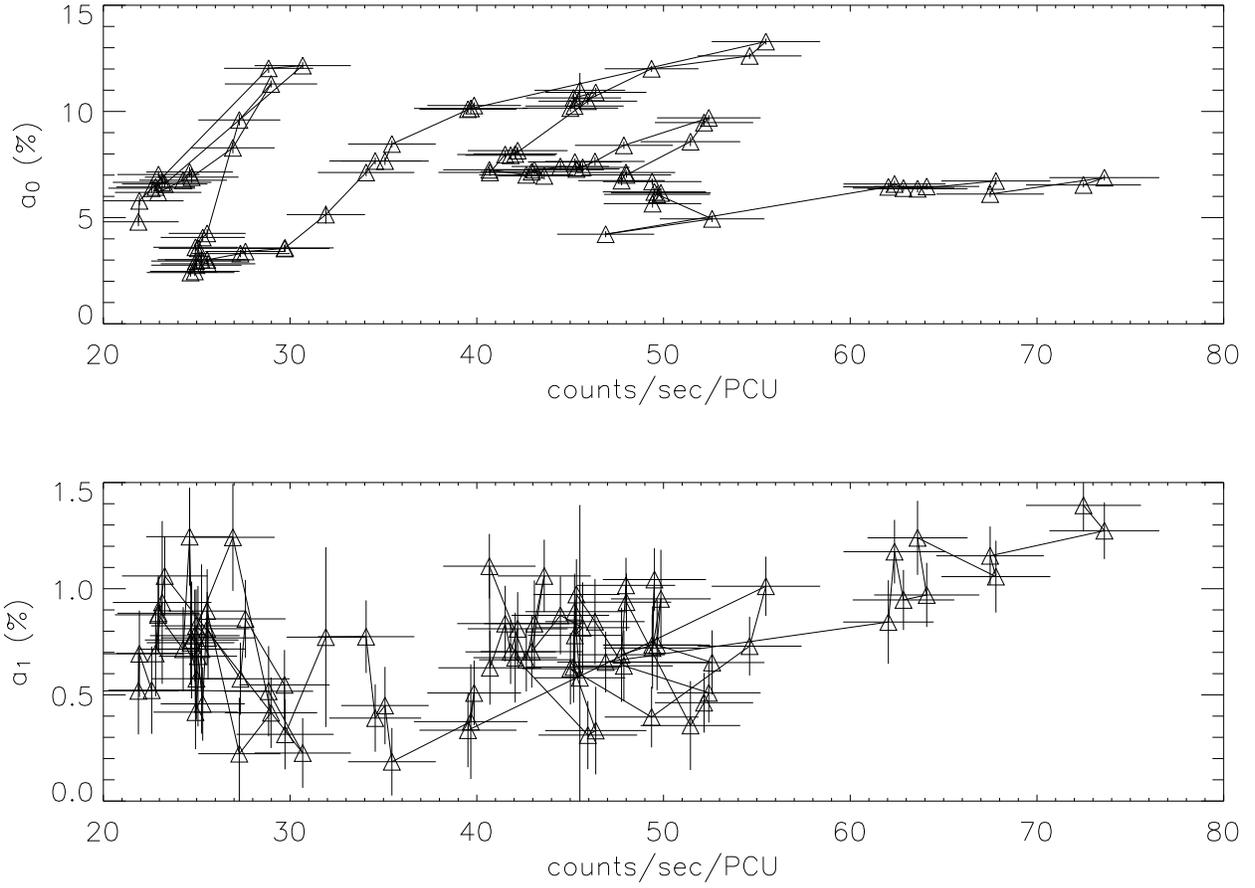} \caption{Best-fitting parameters
of the fractional pulse amplitude $a_0$ (up panel) and the
fractional first harmonic amplitude $a_1$ (bottom panel) with
respect to the averaged net count rate of the pulse profiles.
   }\label{fig_J1807_psfile}

\end{figure}

Figure \ref{fig_J1807_psfile} shows the different behaviors of
$a_0$ and $a_1$ with respect to the averaged count rate of the
pulse profile. Combining the fourth panel of Fig.
\ref{fig_outburst}, we find that in the normal state, the first
harmonic is relatively strong, but it becomes weaker (with $a_1$
decreasing from $1.4\%$ to about $0.9\%$) in the decay of the
outburst, whereas $a_0$ keeps at $\sim 6.5\%$; in the flares,
$a_0$ is positively correlated with the averaged count rate,
whereas $a_1$ is generally in a lower level around $0.7\%$ over
the flares. The variational range of $a_0$ is $\sim 2\% - 14\%$.

\subsection{Aperiodic Timing Behavior}

\begin{figure}[ht]
\epsscale{1.2} \plottwo{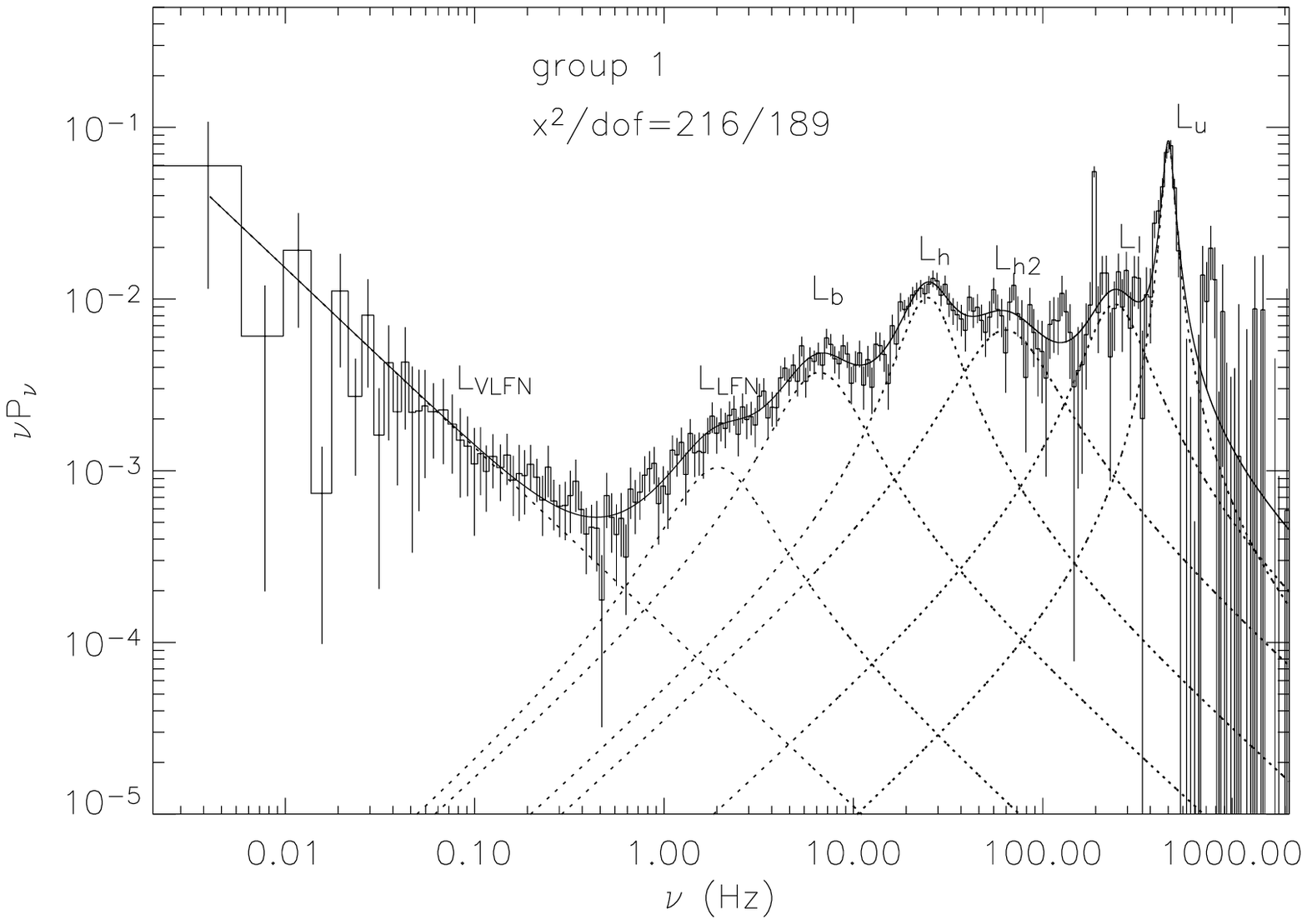}{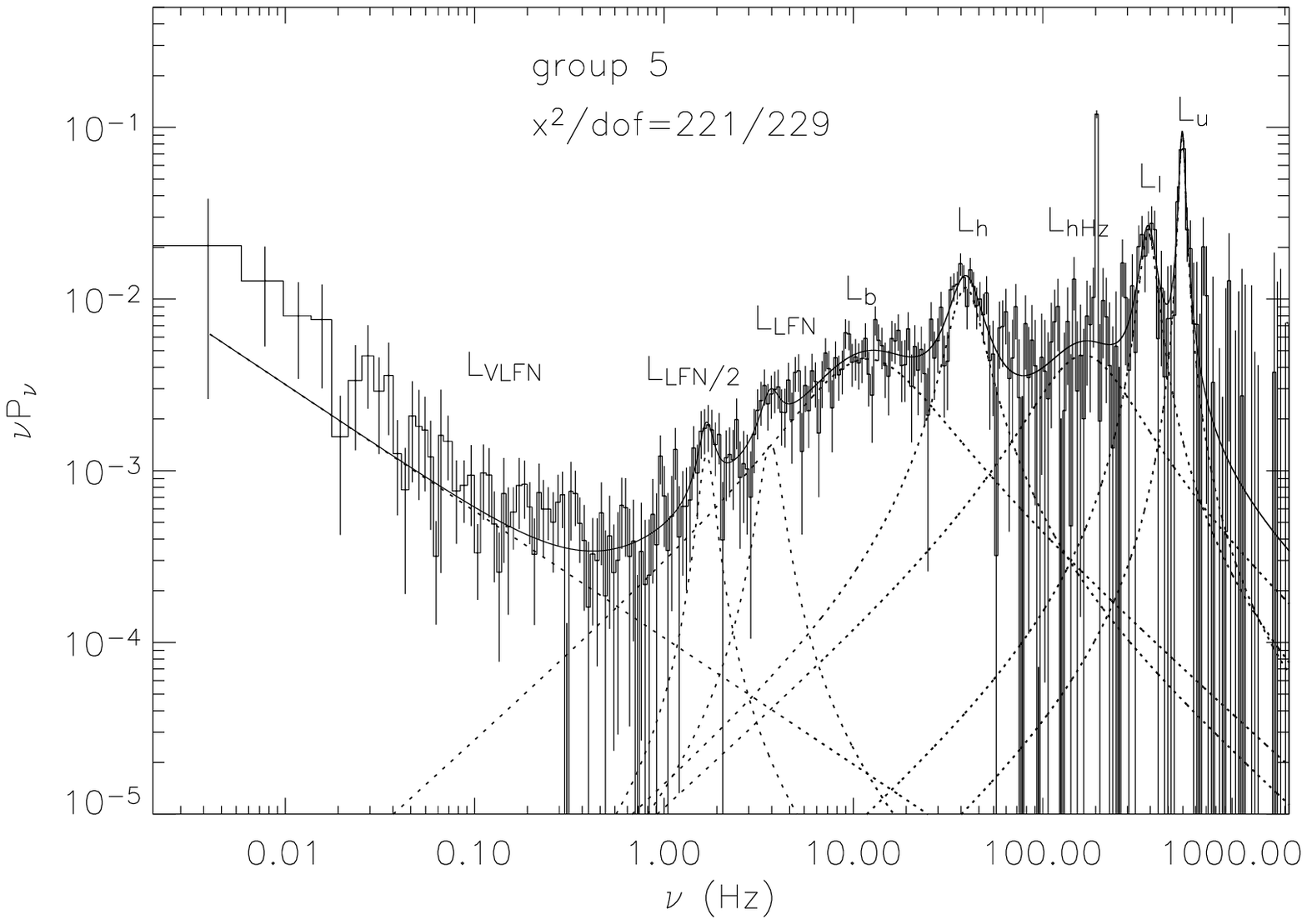} \caption{Power density
spectra in $\nu P_{\nu}$ of group 1 (left) and group 5 (right) for
XTE J1807--294. Solid lines mark the best-fit model, and dotted
lines are the components of every Lorentzian. The components are
identified by their distributions in the frequency-frequency
diagram of Fig. \ref{fig_relation}. }\label{fig_pdsfit}
\end{figure}

We fit the power density spectra by one power-law component plus
three to seven Lorentzian components. Errors on the fit parameters
are determined using $\Delta \chi^2=1$. We list the characteristic
parameters of the Lorentzians, e.g. $\nu_{max}$, $Q$ and $rms$ in
Table 2, the power-law index $\alpha$ of the VLFN and the
$\chi^2/{\rm dof}$ of the fitting in Table 3. In our investigated
episode, kHz QPOs have been detected in groups 1-7 and 9-10, and
undetectable in the other groups on account of the poor statistics
of the power spectra above $\sim$10 Hz. The typical power spectra
of XTE J1807--294 are shown in Fig. \ref{fig_pdsfit}, where group
1 is chosen to stand for the non-flare state and group 5 is taken
as a representation of the flare state.

\begin{figure}[ht]
\epsscale{1.} \plotone{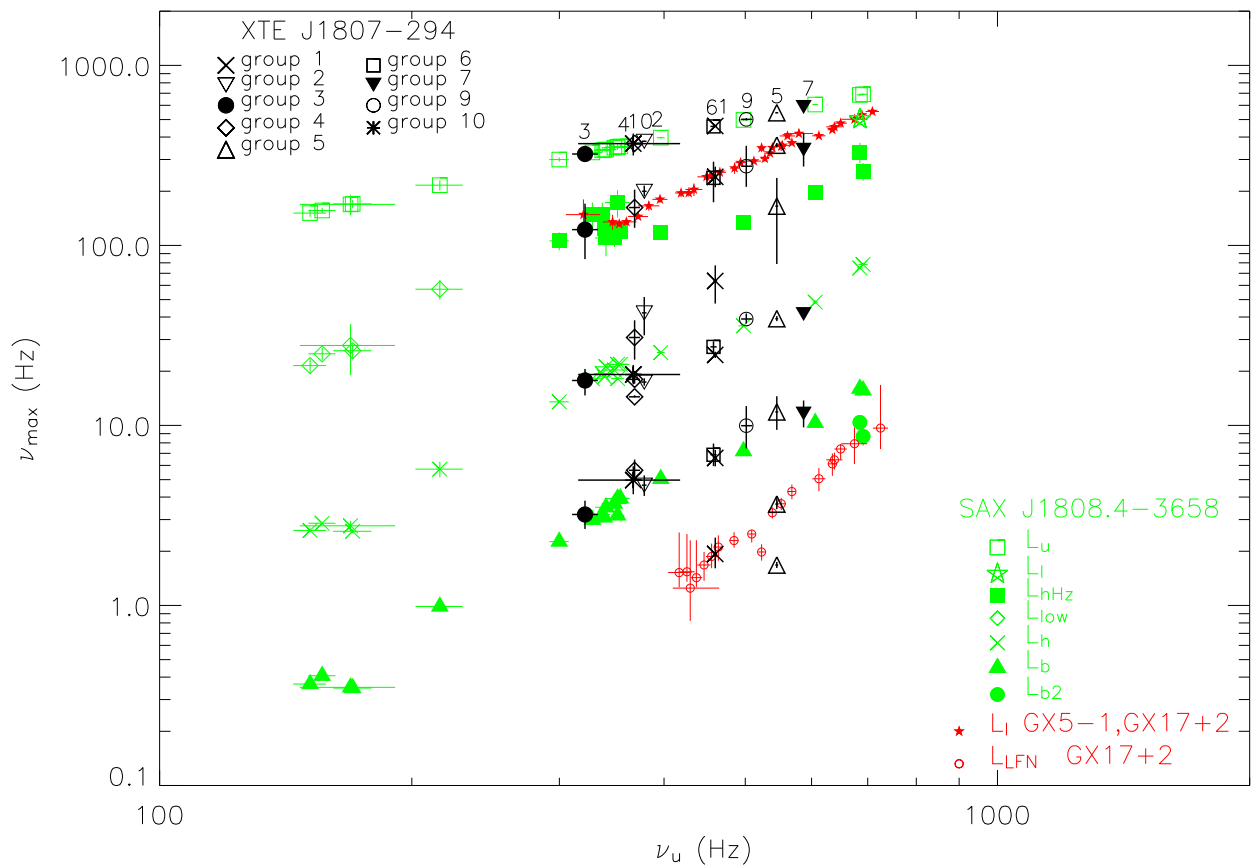} \caption{Characteristic frequencies
of the power spectral components of XTE J1807--294 versus the
upper kHz QPO frequency. The data points of SAX J1808.4--3658
 \citep{vstr05} are drawn for comparisons. Also shown are L$_{\ell}$ of the Z sources
GX 5--1 \citep{jonk02} and GX $17+2$ \citep{homan02}, and
L$_{LFN}$ of GX$17+2$, where the frequencies of these Z sources
are all shifted by a factor of 1/1.5.
 }\label{fig_relation}
\end{figure}

We plot the characteristic frequencies of the Lorentzians versus
the upper kHz QPO frequency in Fig. \ref{fig_relation}. It shows
that XTE J1807--294 has a frequency-frequency correlation similar
to that of SAX J1808.4--3658 \citep{vstr05}. Most of the
Lorentzian components can then be identified in the scheme of
\citet{bpk02} and \citet{vstr03}. In Fig. \ref{fig_relation},
L$_u$ is the upper kHz QPO; L$_\ell$ is the lower kHz QPO;
L$_{hHz}$ is the hectohertz Lorentzian, a broad Lorentzian with a
frequency around 150 Hz; L$_{b}$ is the break Lorentzian, a
band--limited noise component; L$_h$ is the low frequency
Lorentzian just above L$_{b}$. It should be noted that when the
frequency of L$_\ell$ approaches the range of $100\sim 150$ Hz, it
becomes difficult to be distinguished from L$_{hHz}$, especially
when one of them is not detected in the spectra, e.g. groups 3 and
4 (see Fig. \ref{fig_relation}). In Table 2 we put `?' on their
identifications.

\begin{figure}[ht]
\epsscale{1.} \plotone{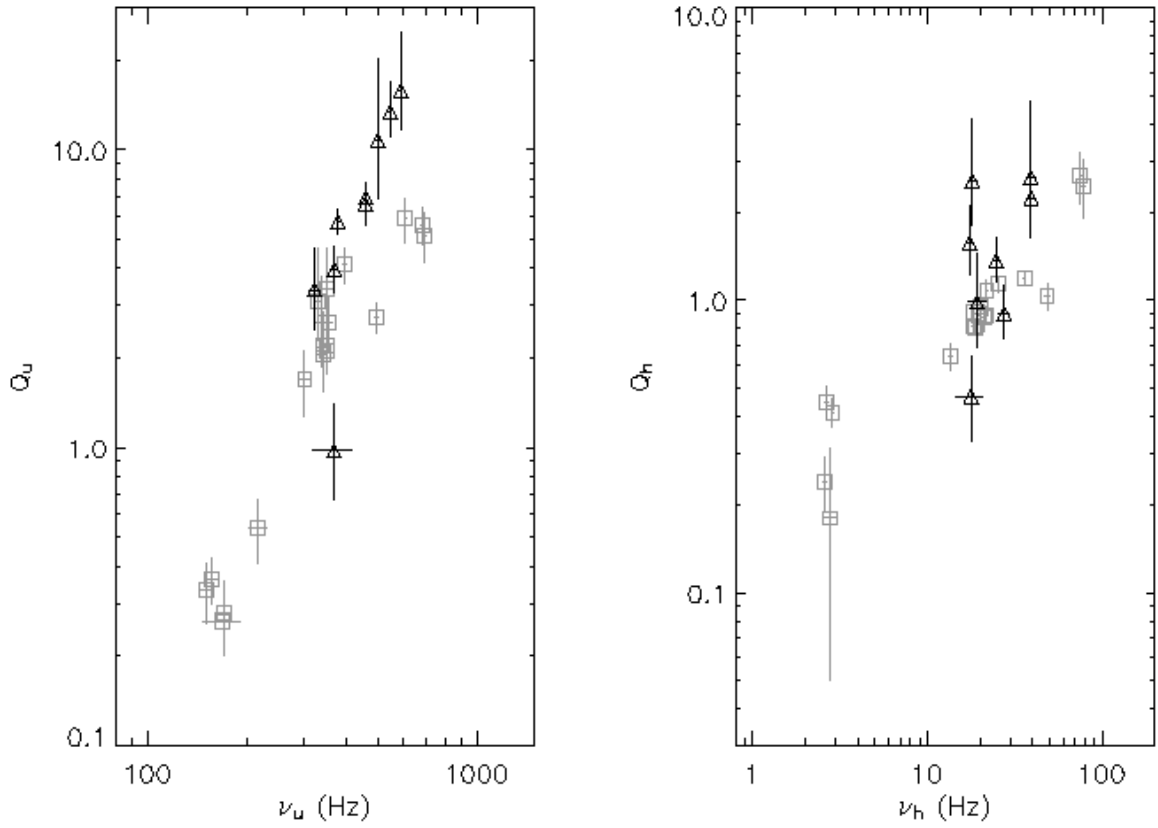} \caption{Q value of L$_{u}$
(left) and L$_{h}$ (right) of XTE J1807--294 (triangles)
 plotted versus the corresponding
characteristic frequencies respectively. Data points of SAX
J1808.4--3658 \citep{vstr05} are plotted by grey squares for
comparisons.} \label{fig_qvsnu}
\end{figure}

In Fig. \ref{fig_relation} we multiply the frequencies of twin kHz
QPOs of Z sources GX 5$-$1 and GX 17$+2$ by a factor of 1/1.5 for
comparison. Their shifted frequencies are almost consistent with
the frequency--frequency distribution of the twin kHz QPOs of XTE
J1807--294. We draw $Q$ versus $\nu_{\rm max}$ for both L$_u$ and
L$_h$ in Fig. \ref{fig_qvsnu}. It proves that the frequency shift
exists in the upper kHz QPO frequencies in XTE J1807--294 as in
the case of SAX J1808.4--3658 \citep{vstr05}.

Except of the five major Lorentizans of L$_u$, L$_\ell$,
L$_{hHz}$, L$_h$ and L$_{b}$, the power spectra of XTE J1807--294
present some additional components (Fig. \ref{fig_relation}).
One of the components is the Lorentizan of L$_{h2}$, with
frequency nearly double of L$_{h}$. If L$_{h}$ corresponds to the
HBO component of Z sources (see van der Klis 2004 for a review),
L$_{h2}$ would correspond to the HBO harmonic. Such harmonic has
never been reported in atoll sources or other accreting
millisecond pulsars. Another additional component is the
Lorentizan of L$_{LFN}$, which takes the position of the shifted
low-frequency noise (LFN) component of GX$17+2$ in the diagram
(Fig. \ref{fig_relation}). The fitted power law index of
$\nu_{LFN}$ vs. $\nu_{u}$ is $3.42\pm 0.29$, larger than the index
of $\nu_{b}$ vs. $\nu_{u}$ of $\sim$2.7. We use 'Ftest' to check
the probability of the components whose errors are less than
$2\sigma$. All of the weak noise components listed in Table. 2
have Ftest probabilities less than $0.03$.

The simultaneous detection of L$_{b}$ and L$_{LFN}$ in XTE
J1807--294 makes it evident that L$_{LFN}$ is a different
component from L$_{b}$. We infer from this finding that the
detected LFN in GX $17+2$ \citep{homan02} should be a different
component from those observed in other Z sources. The shift factor
of about 1.5 in L$_{LFN}$ as well as in L$_u$ and L$_{\ell}$ imply
that these three components could originate from the same inner
disk region.

\subsection{Correlations Between Count Rate, Pulse Profile and kHz QPO }

\begin{figure}[ht]
\epsscale{1.} \plotone{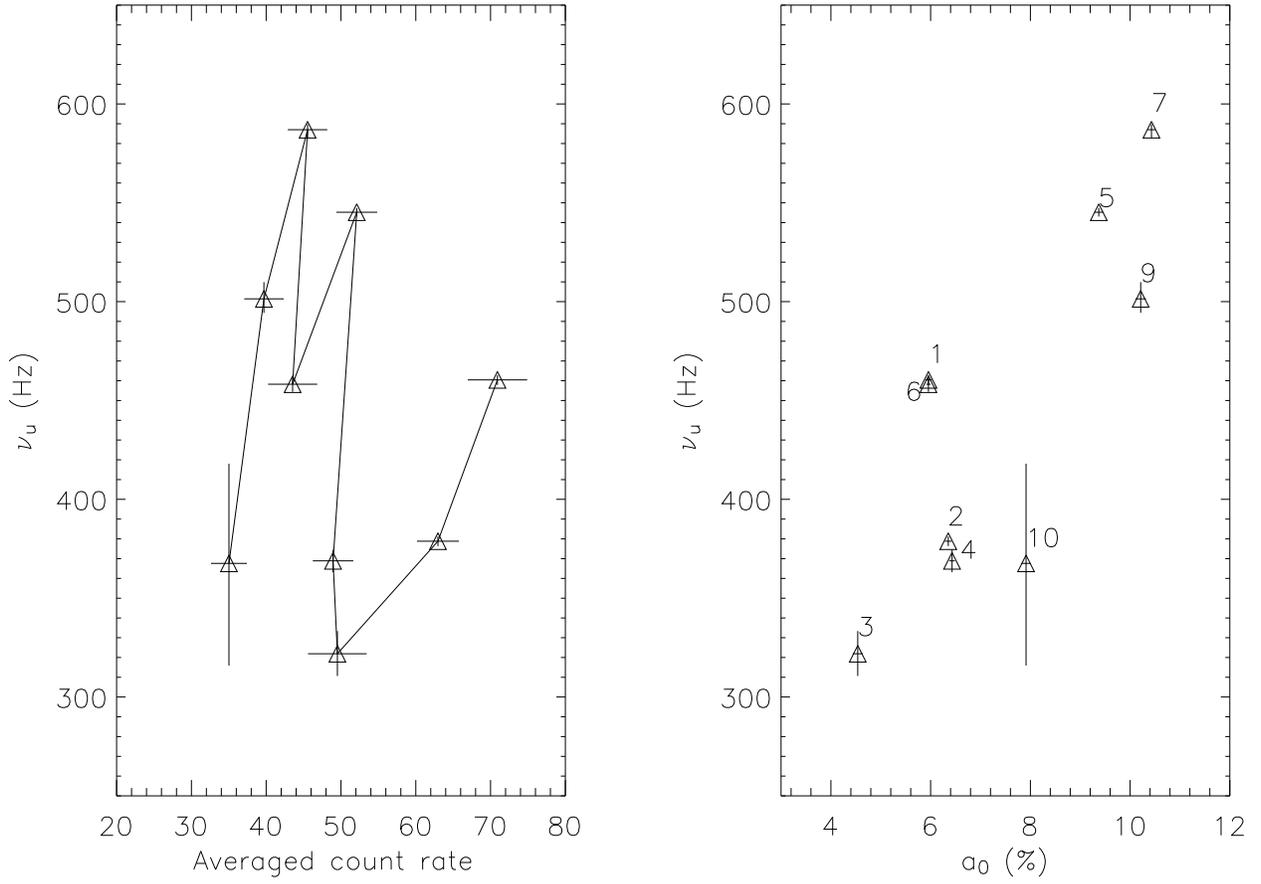} \caption{Characteristic frequencies
of the upper kHz QPOs versus the averaged count rate (left) and
the fractional pulse amplitude $a_0$ (right) of groups 1--7,
9--10. The group numbers are noted in the right panel.}
\label{fig_a0QPO}
\end{figure}

We draw in Fig. \ref{fig_a0QPO} the upper kHz QPO frequency
($\nu_{\rm u}$) versus the averaged count rate and the fractional
pulse amplitude ($a_0$) of the group, respectively. The averaged
count rate and $a_0$ of every group were calculated using the same
methods described in section 3.2.

We can see from Fig. \ref{fig_a0QPO} and Fig. \ref{fig_relation}
that, while the frequencies of the Lorentizans decrease with the
count rate in the non-flare state of groups 1--3 and increase with
flare intensities in groups 4--6 and 7--10, they do not vary
monotonously with the X-ray count rate in the whole episode. The
positive frequency--count rate correlations in the rise of the
flares seem to be steeper than that measured in the normal
outburst decaying state. However, in the whole investigated
episode there exists a noticeable positive $\nu_{\rm u}$ vs. $a_0$
relation with a linear slop coefficient of $\sim 35$ (Fig.
\ref{fig_a0QPO}).

\section{DISCUSSION}

Our motivation to do timing analysis on XTE J1807--294 is to probe
the effects on the neutron-star surface X-ray emissions made by
the disk flow activities, so we focus on analyzing the variations
of pulse profiles and power density spectra during the evolution.

We find several pairs of twin kHz QPOs in the power density
spectra of XTE J1807--294. The frequency separation $\Delta \nu$
ranges from $\sim 179\,\rm{Hz}$ to $\sim 247\,\rm{Hz}$, close to
the spin frequency of $\nu_{\rm s}\sim 191\,\rm{Hz}$. Several kHz
QPO models, considering the coupling of the neutron-star spin to
the Keplerian orbit motion of material in the accretion disk, have
been put forward to explain this specific feature. We notice that
most of these models concern inhomogeneous accretion flow by for
example accreting blobs, density fluctuations in the innermost
part of the accretion disk, or high-density loops confined along a
magnetic field line. In the modified sonic-point beat-frequency
model \citep{lamb01,lamb03}, the accreting blobs are introduced to
explain the deviation of $\Delta \nu$ from the constant value of
the spin frequency by considering their relative motions to the
neutron star. In the resonant disk oscillation models
\citep{kluz04,lee04}, the twin kHz QPOs are explained by a
nonlinear resonance excited by coupling disk oscillation modes to
the neutron star spin. \citet{petr05} proved further that a
rotating misaligned magnetic field of a neutron star or a rotating
non-axially symmetric gravitational potential can provide the
coupling; their perturbations on the accretion disk can give rise
to very pronounced density fluctuations at the inner edge of the
disk. Most recently, \citet{li05} pointed out an alternative
interpretation for the lower kHz QPO in terms of the standing kink
modes of magnetic loop oscillations at the inner edge of the disk,
where the loops with high-density plasma and small cross section
can be produced from the reconnection of the azimuthal magnetic
field lines.

The variation of kHz QPO frequencies can be explained by the
movement of the inner edge of the accretion disk in the radiative
disk truncation model \citep{mill98,klis01}. For example, the
observed positive frequency-count rate correlations on timescales
of hours to days in some atoll sources have been interpreted as
the changes in the inner disk radius due to disk accretion rate
fluctuations. We find a similar frequency-count rate correlation
in XTE J1807--294, but we notice that the shift of the QPO
frequency is related to some broad ``puny" flares. The observed
frequency-count rate correlation in the rise of the flares seems
to be steeper than in the normal state. This result means that (1)
in the case of the same count rate, the kHz QPO frequency measured
in the rise of the flare is larger than that in the normal state;
(2) for a certain kHz QPO frequency range, the variation of the
count rate is smaller in the flare than in the normal state. A
possible explanation is that there are blobs of high mass
densities at the inner edge of the accretion disk and the they can
persist in time scales of hours to days. The formation and the
detailed structure of such a kind of blob are still uncertain. We
refer to the hydrodynamic simulation result by \citet{petr05}
mentioned in the above paragraph for instance. It might be
expected that the relative intensity of the flares to the nearby
normal state could be a measure of the mass density of the blob;
while the limited time intervals of the flares could be related to
the size of the blobs. However, besides the variation of mass
accretion rate due to accreting the blobs, there are other
mechanisms which can influence the X-ray intensity and kHz QPO
frequency, such as nuclear burning on the neutron star surface
\citep{bild93,yu02}. To make a further quantitative investigation
of the blobs from the observation, we have to distinguish their
different effects firstly.

Accreting inhomogeneous flow as one of the possible origins of
flares has been suggested by \citet{moon03} for LMC X--4. In the
``broad flare" of LMC X--4, the spectrum is softened and the pulse
profile becomes simple sinusoids. These features are also observed
in the broad ``puny" flare of XTE J1807--294. In the massive X-ray
binary LMC X--4, the X-ray emission mechanisms of the magnetic
neutron star (with surface magnetic field strength $B\sim
10^{12}\,\rm{G}$) are too complex to be modelled. However, the
spectrum of XTE J1807--294 can be well fitted by an absorbed
black-body plus Comptonization model; thus we can hope to learn
more about the X-ray emission properties by fitting the pulse
profiles of XTE J1807--294 as done by \citet{Gier02,GP05} for SAX
J1808.4--3658 and XTE J1751--305.

Our preliminary results of the pulse-profile analysis present that
both $a_0$ and $a_1$ behave differently in the flares and in the
non-flares. As for the latter case, while the X-ray count rate
decreases, $a_0$ keeps at about $6.5\%$, whereas $a_1$ decreases
from $\sim 1.4\%$ to $\sim 0.9\%$. However, in the former case,
$a_1$ is generally in a low level of $0.7\%$, whereas $a_0$
increases with the flare intensities. The value of $a_0$ spans in
the range of $\sim 2\% - 14\%$, comparable to those observed in
some thermonuclear bursts of XTE J1814--338 \citep{stro03,watt05}
and some non-pulsing LMXBs \citep{muno02}. According to the X-ray
emission model of \citet{pout03}, the variation of the pulse
profile can be determined by the changes of the surface-emission
parameters, such as the hot spot size, position, and even emission
patterns, e.g. the optical depths of the black-body emission and
the Comptonization, respectively. We will put the detailed
spectroscopic and pulse-profile modelling analysis in a future
work. According to this model, the above results have already
shown that the accretion emissions at the neutron star
surface/boundary-layer are very sensitive to the accretion flow
inhomogeneities. The most direct evidence is the observed positive
correlation between kHz QPO frequency and $a_0$ in the
investigated episode.

Although we infer from the above results that the broad ``puny"
flares could be mainly due to accreting blobs, we cannot exclude
the possibility of the existence of some special mode of nuclear
burning on the neutron star surface. \citet{bild93} firstly
proposed that a time-dependent nuclear burning (slow fires) on
patches of the neutron star surface can occur in an intermediate
accretion regime, where the accretion rate is sub-Eddington but is
still too high to allow type I thermonuclear burst, e.g. $5\times
10^{-10}\,\rm{M_{\sun}yr^{-1}}<\dot{M}<10^{-8}\,\rm{M_{\sun}yr^{-1}}$
for a pure helium burning case \citep{bild95}. For XTE J1807--294,
the conditions for slow fires seem to be satisfied qualitatively:
no type I thermonuclear burst has been observed and the source
luminosity drops from $1.3\times 10^{37}\,\rm{erg\,s}^{-1}$ to
about $3.6\times 10^{36}\,\rm{erg\,s}^{-1}$ (assuming the distance
of 8 kpc) in the investigated episode \citep{fala05}. However, the
property of the companion star is still uncertain. It could be a
He dwarf with a low inclination angle ($i<30^{\circ}$) located at
$\sim 8\,\rm{kPc}$, or a C/O dwarf with high inclination angle
($60^{\circ}<i<83^{\circ}$) at about 3 kPc \citep{fala05}, which
makes it difficult to give a quantitative estimation of the
ignition and the spreading of the fires on XTE J1807-294. Even
though, the observations of the low-level luminosity variations on
timescales of $10^4\, \rm{s}$ and the existence of the very low
frequency noise (VLFN) in the mHz--Hz range are all consistent
with the features of the slow fires predicted by the theoretical
investigations of \citet{bild95}. If the slow fires exist, the
position and the size of the black-body radiation on the neutron
star surface will certainly be influenced, making impacts on the
soft-band pulse profiles and time lags. We will consider this
effect in another work.

\section{SUMMARY AND CONCLUSION}

XTE J1807--294 shows significant variations of pulse profiles and
kHz QPO frequencies during the source evolution in March, 2003. It
thus provides a good chance to probe the neutron-star
surface/boundary layer X-ray emissions, the inner disk flow
activities and their correlations.

The main results we get from {\it RXTE} data analysis are related
to some specific ``puny" flares, featured by low-amplitude
intensity fluctuations, hours-to-days long timescale, stronger
soft emissions and significant increase of fractional pulse
amplitude with the flare intensities. A positive kHz QPO
frequency--count rate correlation is also found to be connected
with the flares: the correlation seems to be steeper in the rise
of the flares than in the normal state. However, in the whole
investigated episode, the   kHz QPO frequency is positively
correlated with the fractional pulse amplitude. In the normal
state, the fractional pulse amplitude keeps constant and the first
harmonic content decreases with the outburst decay.

We propose that accreting high-mass-density blobs could be
responsible for all of the above features of the ``puny" flares,
and conclude that neutron-star surface emissions should be very
sensitive to the disk flow inhomogeneities in the hard state.
Since the variation of the fractional pulse amplitude in the flare
is comparable to that in the thermonuclear burst, the effect of
the accretion flow inhomogeneity on material depositions at the
neutron star surface should be carefully considered in the study
of thermonuclear burst oscillations.

Detailed spectroscopic and pulse-profile analysis on XTE
J1807--294 will be done to determine the physical parameters of
the surface/boundary-layer emissions in the whole investigated
episode in a subsequent work.

\acknowledgments

We would like to thank the anonymous referee for his or her very
helpful comments, and thanks are also due to  M. Linares and S.
Zhang for their helpful discussions on the {\it RXTE} data
analysis. This research has made use of the data obtained from the
High Energy Astrophysics Science Archive Research Center provided
by NASA's Goddard Space Flight Center. This work is subsidized by
the Special Funds for Major State Basic Research Projects and by
the National Natural Science Foundation of China.

\clearpage

\clearpage

\clearpage

\clearpage

\clearpage

\clearpage

\clearpage

\clearpage

\begin{table}
\caption{Observational parameters of the 16 groups in Fig.
\ref{fig_outburst}.} \label{tab_observation}
%\vspace{12pt}
\begin{tabular}{clccl}
\hline
No. & Observation ID                                    & Start         & End      & Exposure (s)   \\
\hline
1   & 70134-09-02-00,                                & Feb. 27  11:44:32 & March 1 00:44:08  & 9.712E+03 \\
    & 70134-09-02-01,\\
    & 80145-01-01-01\\
\hline
    & (80145-01-)                                    &                   &                  &            \\
2   & 01-03, 01-04, 01-00                            & March 1  12:44:04 & March 2 03:43:44 & 1.133E+04  \\
3   & 02-00                                          & March 4  22:53:04 & March 5 01:31:44 & 6.032E+03  \\
4   & 02-03, 02-02, 02-01,                           & March 6  16:04:00 & March 8 00:30:40 & 1.912E+04  \\
    & 03-03, 03-02 \\
5   & 03-00                                          & March 8  23:42:08 & March 9 03:19:44 & 8.096E+03  \\
6   & 02-04, 02-06, 01-05,                           & March10  00:28:32 & March13 01:35:44 & 4.984E+04  \\
    & 03-01\\
7   & 02-05, 04-01                                   & March13  18:27:12 & March14 01:44:32 & 1.536E+04  \\
8   & 04-00                                          & March14  21:17:20 & March15 01:55:44 & 9.696E+03  \\
9   & 04-02, 04-10                                   & March16  00:06:08 & March16 04:15:44 & 7.792E+03  \\
10  & 04-03                                          & March16  20:37:20 & March17 00:46:40 & 9.696E+03  \\
11  & 04-04, 04-09, 04-11,                           & March17  18:45:20 & March19 22:14:40 & 1.598E+04 \\
    & 04-05, 04-07\\
12  & 04-08, 05-01, 05-03,                           & March20  20:51:28 & March23 19:26:40 & 2.941E+04 \\
    & 05-02, 05-00, 05-11,\\
    & 05-05\\
13  & 05-06, 05-04, 05-08                            & March25  00:46:24 & March27 00:43:44 & 6.128E+03  \\
14  & 05-07, 05-10, 05-09,                           & March27  16:03:28 & March29 19:29:36 & 1.984E+04  \\
    & 06-00, 06-07, 06-01,\\
    & 06-10\\
15  & 06-02                                          & March30  14:35:28 & March30 16:56:32 & 5.840E+03   \\
16  & 06-03                                          & March31  11:48:32 & March31 17:20:32 & 9.664E+03  \\
\hline
\end{tabular}
\end{table}

\clearpage

\begin{center}
\begin{deluxetable}{llccccccccc}
\tabletypesize{\tiny} \tablewidth{0pt}
\tablecaption{Characteristic parameters of the fitted Lorentzians
of groups 1--10.}\label{tab_pdsfit} \tablehead{ \colhead{No.} &
&\colhead{L$_{LFN/2}$} & \colhead{L$_{LFN}$(L$_{b2}$)} &
\colhead{L$_{b}$} & \colhead{L$_{h1}$} & \colhead{L$_{h}$} &
\colhead{L$_{h2}$} & \colhead{L$_{hHz}$} &
\colhead{L$_\ell$} &  \colhead{L$_u$} \\
}

\startdata

  & $\nu_{\rm max}$(Hz) &          --  & $  1.93_{-0.32}^{+       0.45}$   & $  6.60_{- 0.67}^{+0.71}$    & --                & $ 24.7\pm 1.0$        & $ 64_{-16}^{+     14}$  &  --    & $240_{-30}^{+     33}$   & $460.4\pm 2.5$  \\
1 & $ Q $               &          --  & $  0.72_{-0.28}^{+ 0.40  }$   & $  0.84_{-0.19}^{+0.23}$    & --                & $  1.37\pm 0.25$        & $  0.76_{-0.34}^{+   0.45}$  &   --   & $  1.27_{-0.45}^{+       0.67}$   & $  6.9\pm 0.7$  \\
  & $rms(\%)$           &          --  & $  4.5_{-1.1}^{+       1.3}$   & $  8.0\pm 1.0$    & --                & $ 11.2_{-2.5}^{+       3.3}$        & $ 10.6_{-2.5}^{+       2.9}$  &   --   & $ 10.6_{-2.5}^{+       3.0}$   & $ 13.5_\pm 1.1$  \\

\hline

  & $\nu_{\rm max}$(Hz) &          --  & --              & $  4.66\pm 0.61$     & --                 & $ 17.41_{-0.76}^{+0.86}$        & $ 42.2_{-10.5}^{+     9.5}$ & --            & $200_{-12}^{+   15}$   & $378.8\pm 2.4$  \\
2 & $ Q $               &          --  & --              & $  0.42\pm 0.11$     & --                 & $  1.57_{-0.35}^{+    0.54}$        & $  0.44_{-0.20}^{+  0.25}$ &  --           & $  2.19_{-0.79}^{+     1.48}$   & $  5.74\pm 0.56$  \\
  & $rms(\%)$           &          --  & --              & $  8.9\pm 1.1$     & --                 & $  8.1\pm 1.7$        & $ 14.49_{-2.61}^{+  3.43}$ & --            & $  8.09_{-2.39}^{+     3.04}$   & $ 12.8\pm 1.0$   \\

\hline

  & $\nu_{\rm max}$(Hz) &          --  & --              & $  3.20_{-0.54}^{+    0.62}$     & --                 & $ 17.8_{-3.1}^{+    2.8}$        & -- & $122_{-38}^{+   48}$           &  $-->$ ?   & $322\pm 11$  \\
3 & $ Q $               &          --  & --              & $  0.64_{-0.21}^{+   0.30 }$     & --                 & $  0.47_{-0.14}^{+   0.17 }$        & -- & $  0.98_{-0.64}^{+    0.98 }$           &  $-->$ ?   & $  3.39_{-0.89}^{+   1.31 }$   \\
  & $rms(\%)$           &          --  & --              & $  6.4_{-1.3}^{+    1.4}$     & --                 & $ 14.0\pm 1.9$        & -- & $  9.2_{-3.1}^{+     5.7}$           &  $-->$ ?   & $ 10.7_{-2.3}^{+    2.7}$  \\

\hline

  & $\nu_{\rm max}$(Hz)&   --   & --   &$  5.64_{-0.77}^{+0.83}$  & $ 14.43_{   -0.11 }^{+   0.29}$   & $ 18.0_{   -1.0 }^{+   1.2}$  &$ 30.8\pm 7.6$  & $162_{   -37}^{+  42}$  &  $-->$ ?  & $368.9\pm 5.6$ \\
4 & $ Q $              &   --   & --   &$  0.49_{-0.12}^{+0.15}$  & $ 13.8_{   -6.6 }^{+  34}$   & $  2.61_{   -0.79 }^{+   1.60}$  &$  0.68_{   -0.28 }^{+   0.50}$  & $  0.70_{   -0.29 }^{+   0.49}$  &  $-->$ ? & $  3.96_{   -0.64 }^{+   0.81}$\\
  & $rms(\%)$          &   --   & --   &$  1.9\pm 1.3$  & $  2.8_{   -1.5 }^{+   4.0}$   & $  5.6\pm 1.7$  &$ 10.9_{   -3.0 }^{+   3.3}$  & $ 11.8_{   -2.8 }^{+   3.3}$  &  $-->$ ?  & $ 11.8\pm 1.5$\\

\hline

  & $\nu_{\rm max}$(Hz) &$  1.68\pm 0.06 $   &$  3.66_{ -0.23  }^{+   0.29  }$   &  $ 11.9_{ -2.4  }^{+     2.6  } $  & --       & $ 39.2_{ -1.2  }^{+   1.3  } $       &  --        &$165_{ -86 }^{+ 72 }$  & $  360\pm 11$       & $545.3\pm 2.1$  \\
5 & $ Q $               &$   4.7_{ -1.8 }^{+   3.1  }$   &$  3.6_{ -1.8 }^{+    3.7  }$   &  $  0.42_{ -0.16 }^{+  0.19  } $  & --       & $  2.24_{ -0.46 }^{+    0.63  } $       &  --        &$  0.69_{ -0.52 }^{+  0.85  }$  & $  4.6_{ -1.2 }^{+    1.8  }$       & $ 13.5_{ -2.3 }^{+   3.4  }$  \\
  & $rms(\%)$           &$  2.0_{ -0.7 }^{+     0.9 }$   &$  2.5_{ -1.0 }^{+     1.5 }$   &  $ 11.4_{ -1.6 }^{+   1.9 } $  & --       & $  9.0_{ -1.4 }^{+     1.5 } $       &  --        &$  9.8_{ -3.3 }^{+   4.5 }$  & $  9.1_{ -2.1 }^{+     2.3 }$       & $ 10.4_{ -1.7 }^{+    1.9 }$  \\

\hline

  & $\nu_{\rm max}$(Hz) &  --          &  --             & $   6.87_{ -0.93 }^{+ 1.1 } $    & --       & $ 27.3\pm 2.0$      & --     &  --         &$237_{ -64 }^{+   55 }$    & $458.2_{  -3.8 }^{+4.0 }$  \\
6 & $ Q $               &  --          &  --             & $   0.38_{ -0.09}^{+  0.11 } $    & --       & $  0.90_{ -0.16}^{+    0.22 }$      &  --    &  --         &$  0.57_{ -0.25}^{+   0.31}$    & $  6.6\pm 1.1$   \\
  & $rms(\%)$           &  --          &  --             & $  9.9\pm 1.1 $    & --       & $  11.6\pm 1.3$      &  --    &  --         &$  13.5_{ -2.2 }^{+ 2.3}$    & $  10.1_{ -1.2 }^{+ 1.4}$   \\

\hline

  & $\nu_{\rm max}$(Hz) &  --          &  --             & $ 11.6_{-1.9  }^{+ 2.1  }$  & --      &   --         & $ 41.5_{-2.3  }^{+ 2.5  }$   &  --  & $340_{-65 }^{+ 82 }$    & $587.0_{-4.1  }^{+ 2.8  }$\\
7 & $ Q $               &  --          &  --             & $  0.38\pm 0.12$  & --      &   --         & $  1.77_{-0.43}^{+ 0.60    }$   &  --  & $  1.26_{-0.54}^{+ 1.05    }$    & $ 15.8_{-4.0}^{+ 9.2    }$ \\
  & $rms(\%)$           &  --          &  --             & $ 12.8_{-1.5}^{+  1.8   }$  & --      &   --         & $  9.2_{-1.7}^{+  1.9   }$   &  --  & $ 10.8_{-3.3}^{+  4.0   }$    & $  9.4_{-2.7}^{+  3.8   }$\\

\hline

  & $\nu_{\rm max}$(Hz) &  --          &  --             & $ 10.8_{-1.1 }^{+   1.2}$   & --         & --            & $ 55_{-21}^{+    19}$      &  --         & --            & --        \\
8 & $ Q $               &  --          &  --             & $  0.37_{-0.07}^{+   0.08 }$   & --         & --            & $  1.00_{-0.64}^{+     2.28 }$      &  --         & --            & --        \\
  & $rms(\%)$           &  --          &  --             & $ 14.1_{-1.1}^{+    1.2}$   & --         & --            & $  6.4_{-3.2 }^{+     4.2}$      &  --         & --            & --        \\

\hline

  & $\nu_{\rm max}$(Hz) &  --          &  --             & $  9.9_{-2.5}^{+    2.8}$       & --            & $ 39.0_{-2.2}^{+   2.9}$       & --        &  --           & $275_{-64}^{+ 82}$    &$501.4_{-7.1}^{+   8.5}$  \\
9 & $ Q $               &  --          &  --             & $  0.40_{-0.19}^{+   0.22 }$           & --            & $  2.63_{-0.98}^{+  2.17 }$       & --        &  --           & $  0.79_{-0.36}^{+  0.67 }$    &$ 10.8_{-3.9}^{+  9.6 }$  \\
  & $rms(\%)$           &  --          &  --             & $ 10.7_{-1.9}^{+    2.4}$           & --            & $  6.9_{-2.3}^{+   2.8}$       & --        &  --           & $ 14.9_{-4.2}^{+   5.5}$    &$  8.5_{-3.2}^{+   4.1}$  \\

\hline

  & $\nu_{\rm max}$(Hz) &  --          &  --         & $  4.97_{-0.82 }^{+  0.96  }$  & --             & $ 19.2_{-2.3 }^{+  2.5  }$      & --          &  --           & --         & $368_{-52}^{+  50 }$ \\
10 & $ Q $              &  --          &  --         &   0.00 (fixed)         & --             & $  0.99_{-0.30}^{+     0.45}$      & --          &  --           & --         & $  0.98_{-0.31}^{+     0.42}$ \\
  & $rms(\%)$           &  --          &  --         & $ 12.9_{-1.7}^{+     1.9}$  & --             & $  9.2_{-1.9}^{+     2.1}$      & --          &  --           & --         & $ 16.6_{-3.2}^{+     3.8}$ \\

\hline
\enddata
\tablecomments{In groups 3--4, `$-->$ ?' is used in the case where
 L$_\ell$ can hardly be distinguished from L$_{hHz}$.}
\end{deluxetable}
\end{center}

\begin{center}
\begin{deluxetable}{lccccc}
\tabletypesize{\tiny} \tablewidth{0pt} \tablecaption{Fitted power
law index of the VLFN component together with $\chi^2$ and degree
of the freedom (d.o.f) of the fitting.}\label{tab_pdsfitVLFN}
\tablehead{ \colhead{No.} &\colhead{1} & \colhead{2} & \colhead{3}
&
\colhead{4} & \colhead{5} \\
}

\startdata

$\alpha$ & $-2.04\pm0.16$ & $-1.99\pm0.08$ & $-1.78\pm0.09$ & $-1.86\pm0.14$ &
$-1.74\pm0.18$  \\
\hline

$\chi^2$ & $216$ & $214$ & $208$ & $408$ & $221$ \\
d.o.f & $189$ & $204$ & $195$ & $232$ & $229$ \\
\hline
\hline
 & & & & & \\
No.         & 6 & 7 & 8 & 9 & 10 \\
 & & & & & \\
\hline

$\alpha$ & $-1.90\pm0.06$ & $-1.73\pm0.09$ & $-1.95\pm0.21$ &
$-2.34_{-0.41}^{+0.36}$ & $-1.94\pm0.15$ \\
\hline
$\chi^2$ & $397$ & $289$ & $261$ & $343$ & $267$ \\
d.o.f  & $228$ & $238$ & $226$ & $203$ & $239$   \\
\hline
\enddata

\end{deluxetable}
\end{center}

\end{document}